# Understanding the Tenets of Agile Software Engineering: Lecturing, Exploration and Critical Thinking

Shvetha Soundararajan, James D. Arthur and Amine Chigani

Department of Computer Science
Virginia Tech
Blacksburg, VA 24061 USA
{shvetha, arthur, achigani}@vt.edu

*Abstract –  The use of agile principles and practices in software development is becoming a powerful force in today's workplace. In our quest to develop better products, therefore, it is imperative that we strive to learn and understand the application of Agile methods, principles and techniques to the software development enterprise. Unfortunately, in many educational institutions courses and projects that emphasize Agile Software Development are minimal. At best, students have only limited exposure to the agile philosophy, principles and practices at the graduate and undergraduate levels of education. In an effort to address this concern, we offered a graduate-level course entitled "Agile Software Engineering" in the Department of Computer Science at Virginia Tech in Fall 2009. The primary objectives of the class were to introduce the values, principles and practices underlying the agile philosophy, and to do so in an atmosphere that encourages debate and critical thinking. The course was designed around three central components: (1) teaching the essentials of how one develops a product within an Agile framework, (2) having invited presentation by notable industry experts, and (3) having students  present and discuss current research topics and issues. This paper describes our experiences during the offering of that course, and in particular, the unique perspectives of the class instructor, the teaching assistant and a student who was enrolled in the class.*

*Key words: Agile Instruction, Education, Exploration, Agile Software Development, Principles and Practices of Agility*

## I. INTRODUCTION

We, the members of the Department of Computer Science at Virginia Tech, are dedicated to grooming the next generation of Software Engineers. Hence, the department offers courses representing core areas in Software Engineering and current research foci to better prepare our students for success in the Software Engineering arena.  With Agile Software Development becoming more widely used, we realize that instruction and exploration in Agile within an academic setting is essential.  Hence, in Fall 2009 we offered a 6000-level graduate course entitled "Agile Software Engineering."

### A. Course Objectives

Instruction and exploration are integral components of an effective learning process. We designed our course, therefore, to place equal importance on both, and in particular, while addressing the many facets of Agile Software Engineering. The principal objectives that guided course development were:

1. To introduce the agile philosophy, values, principles and practices
2. To develop a firm understanding of what constitutes agility, and how and why it differs from conventional Software Engineering
3. To think critically about the benefits and limitations in application of agile principles and practices.

The first objective focuses on the motivation, rationale, and fundamentals underlying the Agile approach.  The second objective emphasizes a more in-depth study of Agility and the rudiments of critical thinking within an Agile framework. Finally, the third objective stresses exploration and reasoning about the *raison d'être* of the Agile approach to software development.

### B. Course Design

The 6000-level Agile Software Engineering class was designed for graduate students who already understood conventional software engineering models and practices.  To our delight, more than half of the students had been or were currently working in industry.  The course was designed around three principal components: (1) the essentials of how one develops a product within an Agile framework, (2) in-class presentations by industry experts, and (3) presentations by students to showcase current research topics and issues. The former two components focused more on instruction and the latter on exploration. During all three of the above, discussion, questions and debate were *highly encouraged*. In addition to the presentations, the students were also assigned research papers and in-class "mini-exams".

Table 1 outlines additional class details, .e.g., pre-requisites, enrollment, background knowledge of the students, etc.

TABLE I
CLASS DETAILS

| *Pre-requisites* | <ul><li>Software Engineering (CS 5704) or</li><li>Software Design and Quality (CS 5774) or</li><li>Comparable Software Engineering experience</li></ul> | |
|---|---|---|
| *Number of students at the masters and doctoral degree level* | Ph.D. 8 | Masters 2 |
| *Number of students with industry work experience* | 7 | |
| *Number of students with prior experience in agile development* | 2 | |
| *Guiding Manuscript* | "Agile Software Development Methods: Review and Analysis", Pekka Abrahamsson, Outi Salo, Jussi Ronkainen & Juhani Warsta, ESPOO 2002, VTT Pub | |

The remainder of this paper is organized as follows. Section 2 of this paper provides an overview of experiences from the perspectives of the instructor, the teaching assistant and a student who took the course. Section 3 summarizes students' responses to prominent issues that emerged during class discussions. To encourage critical thinking and exploration, student responses had to be prepared in the form of research papers. Finally, Section 4 summarizes our efforts and perceptions.

## II. EXPERIENCES

### A. Instructor's Perspective

The overarching objective of the class was to introduce the values, principles and practices underlying the Agile philosophy, and to do so in an atmosphere that encourages debate and critical thinking.

*The Basics*: The goal for the initial portion of the class was to introduce the students to the tenets of Agile. More specifically, we wanted the student to (a) develop an understanding of what comprised the Agile philosophy and software development approach, (b) gain an appreciation as to why it is needed, and (c) evolve a questioning mindset that encouraged critical thinking. The latter was particularly important because we wanted the students to develop their own perceptions as to how Agile differed from the more conventional software development models and processes. We began by introducing the Agile Manifesto and discussing the values it espoused. From this point, I provided presentations outlining SCRUM, XP and the Lean Methods. For each, the underlying principles, practices and activities were introduced. Throughout the presentations, questions were asked, and discussions ensued, that focused on comparisons among the methods, and on their similarities and differences with the more conventional development methodologies. To further immerse the students in the "ways of Agile," each was required to identify, investigate and present an additional aspect of the agile software development approach. It is my opinion that while the presentations provided foundational knowledge, the discussions, debates, and student presentations were the tools that began cultivating their (the students) aptitude for critical thinking.

*Invited Speakers*: Three speakers from industry were invited to present and discuss their experiences with Agile software development. The first speaker was Keith Lang who manages the development of Vanguard's Investment Product website. Of particular interest to the students was how they (the Vanguard Agile development teams) were able to have four concurrent agile development efforts, all of which were synchronized to produce a single product. The second speaker was Jason Lee who works for Meridium Corporation, and in particular, has helped integrate their agile practices with those practices fundamental to usability engineering. A substantial amount of discussion revolved around how two seemingly independent processes could be integrated into a single set of complementary practices. The students were skeptical, but Mr. Lee was convincing. The third invited speaker was Dr. Ahmed Sidky, an Agile Consultant at TenPearls. Having consulted around the world, Dr. Sidky introduced the class to cultural issues and how they can and do impact the Agile Coaching process. His presentation and insights provided a glimpse of Agility rarely found in books or papers. Finally, we were fortunate to have Dr. Todd Stevens "attend" many classes, and provide his comments on the topic "*de jour*." Dr. Stevens, a Blacksburg resident, is also a Meridium employee and helps manage their Agile development processes. Throughout the semester, he would provide a "real world" perspective on issues and questions, and often challenged the students to think more critically.

From a pedagogical perspective, these people reinforced the theory of Agility through discussions of actual industry practices. The students expressed their appreciation of this fact in the many complimentary comments that followed.

*Research Presentations*: As a third component of the class, students were required to identify, present and critique published research papers focusing on some critical aspect of Agility. The Research Presentations helped expand the students' perspectives as to what are the important issues facing the Agile community, and forced them (or provided them with the opportunity) to *analyze* current research efforts, *critique* the results, and then *defend* their assessment

*Research Papers:* On a final note, throughout the semester, when (selected) questions arose for which there was no apparent answer, the students were asked to compose a 3-4 page research paper that responded to each question. Part of that task was to identify and cite supporting documents that confirmed their reasoning. This, too, was an activity intended to broaden each student's understanding of the Agile philosophy, and to encourage critical thought.

### B. Teaching Assistant's Perspective

As a graduate student researcher in the field of Agile Software Engineering, it was a great opportunity for me to serve as the Teaching Assistant (TA) for the 6000-level graduate "Agile Software Engineering" class in Fall 2009. My responsibilities as the TA included (1) instruction, (2) assisting with grading assignments, and (3) facilitating and leading class discussions. I had the opportunity to deliver two lectures introducing the agile philosophy, its principles and practices and providing an overview of the existing agile methods. Additionally, I facilitated class discussions and offered insights and opinions about topics including, but not limited to, agile methods, current research trends, and issues faced by the agile community. I presented my current research topic, Measuring Agility, to the class and obtained valuable feedback that has helped immensely in my research progress.

I strongly believe in the agile philosophy. When I started my research in Agile Software Engineering three years ago, I embraced the agile philosophy almost immediately. I expected the same from my fellow graduate students enrolled in the course. That I was mistaken is an understatement. Most of them were reluctant to accept the notion of agility. Though more than half of the number of students in the class had prior Software Engineering industry experience, only two of them had worked for organizations that had transitioned to and agile environment. This contributed significantly to their skepticism. For example, the students found the concept of minimal documentation hard to believe and accept. The class discussions at the beginning of the semester were almost confrontational. More than once, the whole class period was consumed by the discussions that would leave no time for the instructor to proceed with his topic of the day. As the semester progressed, however, the students began to appreciate and embrace the agile philosophy. I could observe that the discussions became more complimentary. The transition was reflected not only in the class discussions but also in their answers to questions in exam questions and assignments. We had the privilege of listening to Dr. Ahmed Sidky who provided us with a presentation outlining his experiences in the agile world. In his talk, he mentioned that the people of an organization ready to transition to agile usually are skeptical about to start with, and more often than not, would only embrace the agile philosophy gradually. It was surprising to see the same trend in an academic setting.

At Tech students are required to evaluate their course instructors and teaching assistants at the end of each semester. An analysis of the student responses revealed that they found my insights valuable. What gives me greater pleasure than positive student evaluation results is that the students now embrace agility and have begun applying the agile philosophy to their own work.

To conclude, I would say that the course was valuable to everyone involved. I believe that courses such as these enrich a student's learning experience. I am looking forward to future offerings of the course.

*C. Student's Perspective*

I am a graduate student member of the software engineering research group. My area of interest deals with architecting large-scale systems, which require process-oriented, plan-driven approaches in their development.

My goal of taking the course was to learn the different methodologies that agileists prescribe to develop software. However, I came into the course with two presumptions. First, I was skeptical that agile methods can scale to accommodate the needs of large-scale systems. Second, I was unsure how agileists claim to develop quality software without architecting for these qualities early on in the life cycle.

The overall experience from the course was both challenging and rewarding at the same time. The main challenge I faced was being unable to reconcile my training and experience in conventional software engineering with the practices of agile methodologies. More specifically, one of the issues that kept coming up in many discussions was that of scalability. Being aware of the complexity of large-scale software development efforts, which is often characterized by multi-team, multi-system, and multi-year development aspects, it was problematic for me to see the applicability of the simplistic agile approaches of small teams, non-emphasis on documentation, and short life cycles in a large development effort.

In spite of this challenge, the experience in the course was stimulating in many ways. The course proved to be an excellent opportunity to broaden my understanding and increase my appreciation of agile methodologies.

First, this course was structured to engage our analytical thinking to explore the opportunities and the limitations of agile methods by focusing on a particular area of agile and preparing a paper and a presentation about it. This component of the course allowed me to reconcile my architecture experience with some of the agile practices and incorporate these new findings into my own work. Such insights, I believe, should help the architecture community a great deal. This aspect of the course also encouraged me to write a short paper about agile and architecting to present in an architecture practitioners' conference.

The second reason why this course was a successful experience is the fact that it changed my perspective about agile. In fact, I do consider myself an agile practitioner since I apply basic agile values and principles in my own work. It has become clear to me that agile is more a way of thinking rather than a set of practices. Shaping my way of thinking about agile helped me incorporate several of its practices into my architecture work that focuses on plan-driven development.

Finally, my experience in this course gave me the chance to realize that although people from both sides of the fence see software development form different prisms, the two approaches can be complimentary as evidenced by the effect this course had on my own work. It also made me realize that the potential of agile methods is promising. Therefore, incorporating agile methods in the computing curriculum, by offering courses such as this one, is fundamental to the success of agile methods.

## III. NOURISHING CRITICAL AND CREATIVE THINKING

As mentioned previously, class discussions formed an integral part of the course throughout the semester. The students were encouraged to critique agile methods, analyze their current state, discuss research issues, suggest solutions etc. During these discussions, some issues were brought to the forefront time and time again. These were important issues that the students knew were addressed in the conventional Software Engineering approaches. So the discussions were always about "*How are these issues addressed in the agile approach?*" In order to urge on the students' critical thinking, we selected three of the major issues and assigned them as research papers. The students were expected to write a 3-4 page paper addressing the issue. Critical thinking also enabled them to be creative in their solution approaches. In the following paragraphs, we discuss the assigned questions, the underlying issues, the students' thought process in outlining solutions, and our observations.

Question 1: Government and contracting organizations expect Agile organizations to meet the same CMMI measurement standards as organizations employing a plan-driven development process. If we wanted a CMMI-like approach to measuring the capability of an Agile organization, what Key Process Areas (KPAs) and measurement indicators would you define?

*Issue*: CMMI focuses on assessing an organization's process and process artifacts. On the other hand, Agile Software Development places emphasis on assessing the working software or product.

*Students' thought process*: The students understood the essence of CMMI. They performed a comparative analysis of CMMI and the agile philosophy. Ways to strike a balance between the two were described.

*Observations:* We identified three approaches that the students suggested in order to nurture a symbiotic relationship between the agile philosophy and CMMI. They are outlined below:

- Identify KPAs from CMMI that are relevant to the agile environment, .e.g., the Project Planning KPA. Modify the assessment criteria and the expected artifacts such that agility is not compromised.
- Attempting to force-fit agile to CMMI, or watering down CMMI to suit an agile environment, is not an effective solution. Develop a completely new set of KPAs that are modeled along the lines of the KPAs in CMMI, but that reflect the agile approach to Software Engineering.
- In analyzing the advantages and disadvantages of both CMMI and agile methods, it is imperative to develop a hybrid approach that provides the best of both worlds. More specifically, we should develop new KPAs and also modify existing KPAs that are suitable to an agile environment.

Question 2: How do we scale agile methods to fit larger scale systems development? Or more specifically, what are some of the issues that would have to be addressed to achieve scalability?

*Issue:* Agile methods were intended for small-scale systems.

*Students' thought process:* For larger scale systems, the students had a mental model of a conventional Software Engineering approach like the waterfall model. They compared the agile approach to software engineering to the conventional approach and identified areas that they thought were crucial to scalability.

*Observations:* The students most often identified the need for comprehensive documentation, team size and system complexity as being at odds with existing agile practices. They also suggested ways to scale the identified factors.

Question 3: To what extent do Agile Methodologies address the maintenance activity?

*Issue:* Very little information is revealed concerning to what extent the Agile Methodologies address the post development maintenance activity.

*Students' thought process:* In conventional software engineering approaches, maintenance is portrayed as a downstream development activity. Some of the students perceived maintenance in this conventional sense. Others envisioned maintenance as a part of the agile development process itself.

*Observations:* The students focused primarily the two approaches given below:

- Maintenance is integrated within the Agile Software Development process. Certain artifacts developed during the process assists with maintenance within Agile Software Engineering.
- Maintenance should be considered as a downstream development activity. After the product is developed, maintenance on that product should be supported.

## IV. CONCLUSION

The intent of the course was to provide an understanding of Agile Software Engineering by studying the basics, questioning the principles and practices, and formulating reasoned opinions through critical thinking. We designed the course to include research presentations, talks by industry experts, class discussions and research paper assignments. We could observe the evolution of critical thought as the semester progressed.

The students were required to complete the end-semester student evaluation surveys. Nine out of the ten students in the class completed the surveys. They all indicated that the course objectives were met. They also found the class discussions and research papers to have been the most valuable components of the class.